\documentclass[12pt]{article}

\usepackage{amssymb}

\newcommand{\BEQ}{\begin{equation}}
\newcommand{\EEQ}{\end{equation}}
\newcommand{\BEQA}{\begin{eqnarray}}
\newcommand{\EEQA}{\end{eqnarray}}
\newcommand{\BEQS}{\begin{displaymath}}
\newcommand{\EEQS}{\end{displaymath}}
\newcommand{\BEQAS}{\begin{eqnarray*}}
\newcommand{\EEQAS}{\end{eqnarray*}}
\newcommand{\BEI}{\begin{itemize}}
\newcommand{\EEI}{\end{itemize}}
\newcommand{\BED}{\begin{description}}
\newcommand{\EED}{\end{description}}
\newcommand{\BEN}{\begin{enumerate}}
\newcommand{\EEN}{\end{enumerate}}

\newcommand{\Z}{\mathbb{Z}}

\newcommand{\R}{\mathbb{R}}
\newcommand{\C}{\mathbb{C}}

\newcommand{\Lieg}{\mathfrak{g}}

\newcommand{\Tr}{{\rm Tr}}

\newcommand{\Exp}{{\rm Exp}}

\title{Zeta-Functions and Star-Products}
\author{\large Frank Antonsen \\University of Copenhagen \\Niels Bohr 
Institute
}

\begin{document}
\maketitle

\begin{abstract}
We use the definition of a star (or Moyal or twisted) 
product to give a phasespace definition
of the $\zeta$-function. This allows us to derive new closed expressions for
the coefficients of the heat kernel in an asymptotic expansion for operators
of the form $\alpha p^2+v(q)$. For the particular case of the harmonic
oscillator we furthermore find a closed form for the Green's function.
We also find a relationship between star exponentials, path integrals and 
Wigner functions, which in a simple example gives a relation between the
star exponential of the Chern-Simons action and knot invariants.
\end{abstract}

\section{Introduction}
The star product on a given symplectic manifold is a very powerful way to
quantise a given classical theory. Consider a symplectic manifold $\Gamma$, 
and let ${\cal A}_0=C^\infty(\Gamma)$ be the Poisson-Lie algebra of 
classical observables. We then define a new algebra ${\cal A}_\hbar$ which is
the quantum analogue of ${\cal A}_0$, by letting
\BEQ
	{\cal A}_\hbar = {\cal A}_o\otimes \C[[\hbar^{-1},\hbar]]
\EEQ
and introducing a star product
\BEQ
	f*g = fg + O(\hbar), f*g- g*f = O(\hbar)
\EEQ
and the Moyal bracket
\BEQ
	[f,g]_M = f*g-g*f
\EEQ
which we demand is a deformation of the Poisson bracket, i.e.,
\BEQ
	[f,g]_M = i\hbar \{f,g\}_{\rm PB} +O(\hbar^2)
\EEQ
The Moyal bracket is also required to be a Lie bracket, thus making ${\cal A
}_\hbar$ into a Poisson-Lie algebra. It can be proven that twisted products
exist on any symplectic manifold, \cite{exists}, and, moreover, that 
different choices of star product corresponds to different choices of
operator orderings, \cite{opord}. Furthermore, a $W_\infty$ symmetry
exists relating the various choices of star product, \cite{winf}, and each
of the various choices specify a characteristic class in the de Rham cohomology
of $\Gamma$.\\
For $\Gamma=\R^{2n}$ the unique star product is
\BEQ
	f*g = fe^{\frac{1}{2}i\hbar \{\cdot,\cdot\}_{\rm PB}} g \equiv
	fg+\frac{1}{2}i\hbar\{f,g\}_{\rm PB}+O(\hbar^2)
\EEQ
leading to the standard Moyal bracket
\BEQ
	[f,g]_M = 2if\sin\left(\frac{1}{2}\hbar\{\cdot,\cdot\}_{\rm PB}\right)g
\EEQ
Thus
\BEQ
	[f,g]_M = i\sum_{n=0}^\infty (-1)^n\frac{(2n+1)!}4^{-n}\hbar^{2n+1}
	\sum_{k=0}^n(-1)^k\left(\begin{array}{c}n\\k\end{array}\right)
	\frac{\partial^nf}{\partial q^{n-k}\partial p^k}
	\frac{\partial^ng}{\partial q^k\partial p^{n-k}}
\EEQ
The star product comes from the Weyl map which associates a function on 
phasespace to each operator $A$ on $L^2(\R^n)$,
\BEQ
	A \mapsto A_W(q,p) \equiv \int e^{iup-ivq}\Tr(\Pi(u,v)A)~dudv
\EEQ
where
\BEQ
	\Pi(u,v) = e^{iu\hat{p}-iv\hat{q}}
\EEQ
is a translation operator on phasespace giving a ray-representation of the
Euclidean group $E_{2n}$, see \cite{rayrep}. The star product and the
Moyal bracket can then be obtained from
\BEQA
	A_W*B_W &=& (AB)_W\\
	\left[A_W,B_W\right]_M &=& (AB-BA)_W
\EEQA
The inverse of the Weyl map associates an operator $f^W$ to each
function $f$ on phasespace
\BEQ
	f^W = \int e^{-iup+ivq}\Pi(u,v)f(q,p)dudvdqdp
\EEQ
and we can then write
\BEQ
	f*g = (f^Wg^W)_W
\EEQ
The Weyl map of the density operator $\rho$ is referred to as the Wigner
function. This formalism can be extended to phasespaces which are not just
$\R^{2n}$, see \cite{ext}.

\section{Star Exponentials and Zeta Functions}
Now, consider a pure state $\rho=|\psi\rangle\langle\psi|$, we then get the
standard formula for the Wigner function, $W=\rho_W$,
\BEQA
	W(q,p)&:=&\int e^{iup-ivq}\Tr\Pi(u,v)\rho dudv\\
	&=& \int \langle\psi(q+y/2)|\psi(q-y/2)\rangle e^{-iyp} \frac{dy}
	{(2\pi)^n}
\EEQA
Let $|\psi_\lambda\rangle=|\lambda\rangle$ be a complete set of eigenstates
to some Hermitian operator $H$. We can then form the heat kernel of $H$
\BEQ
	G_H(q,q';\sigma) = \langle q|e^{-H\sigma}|q'\rangle = \langle q|
	\hat{G}_H|q'\rangle
\EEQ
and as is well known, $G_H$ determines the zeta function of $H$,
\BEQ
	\zeta_H(s) = \frac{1}{\Gamma(s)} \int d\sigma \sigma^{s-1}\Tr \hat{G}_H
\EEQ
This provides us with a point of contact with the Wigner-Weyl-Moyal formalism,
since,
\BEQAS
	\Tr A &=& \int A_W(q,p) dqdp\\
	\left(e^A\right)_W(q,p) &=& \Exp(A_W(q,p))
\EEQAS
where $\Exp$ is the $*$-exponential,\cite{starexp},
\BEQ
	\Exp(f) := \sum_{n=0}^\infty \frac{1}{n!} \underbrace{f*f*...*f}_n
	=\sum_{n=0}^\infty \frac{1}{n!}f^{*n}
\EEQ
Thus
\BEQA
	\zeta_H(s) &=& \frac{1}{\Gamma(s)}\int d\sigma \sigma^{s-1} \Tr 
	e^{-H\sigma}\nonumber\\
	&=& \frac{1}{\Gamma(s)}\int d\sigma \sigma^{s-1}\int \Exp(-H_W(q,p)
	\sigma) dqdp
\EEQA
We want to use this formula to find some new expressions for the 
Schwinger-DeWitt asymptotic expansion of the heat kernel. Suppose
$H=\alpha p^2+...$ we can then write
\BEQ
	G_H(q,q';\sigma) = \sum_{n=0}^\infty e^{-\frac{(q-q')^2}{4\sigma}}
	a_n(q,q') \sigma^{n-d/2}
\EEQ
in $d$ dimensions. This implies
\BEQ
	\int G_H(q,q;\sigma) dq = \sum_{n=0}^\infty \int a_n(q)\sigma^{n-d/2}
	dq = \int \Exp(-H_W(q,p)\sigma) dqdp
\EEQ
Now, when $H=\alpha p^2+...$ then $H_W=\alpha p^2+...$ where the ellipsis $...$
stands for terms which are at most linear in $p$. We can then perform the
integration over $p$ (it is a simple Gaussian integral) to obtain
\BEQ
	\int \Exp(-(\alpha p^2+...)\sigma) dp = (\det\alpha)^{-1/2} \pi^{d/2}
	\sigma^{-d/2} E(q;\sigma)
\EEQ
where $E(q;\sigma)$ is some function depending on the precise form for $H$.
We thus see that the $\sigma^{-d/2}$ factor comes from a Gaussian integration
in this phasespace formalism. Taylor expanding $E(q;\sigma)$ in $\sigma$
will then give us $a_q$ in the Schwinger-DeWitt expansion. Notice that such
asymptotic expansions are rather easily derived with this formalism. For 
instance, if $H=\alpha p^\gamma+v(q)$ then we get a factor $\sigma^{-d/\gamma}$
for $\gamma >1$ from the integration over $p$, and we can then find a 
generalised Schwinger-DeWitt expansion for such (pseudo-)differential 
operators. \\
Also interesting is an expansion in powers of $\hbar$. Suppose $f$ is
independent of $\hbar$, we then write
\BEQ
	\Exp(f) = \sum_{n=0}^\infty \hbar^{2n}{\cal E}_{2n}(f)
\EEQ
By writing down the definition of the twisted product, we see that only
even powers of $\hbar$ can occur -- odd powers only become relevant if
the exponent itself is $\hbar$-dependent. Furthermore, the $\hbar^2$ term,
say, will receive contributions form all the twisted powers of $f$. More
precisely, we see
\BEQ
	f^{*n}=f^n-\frac{\hbar^2}{8}\sum_{k=0}^{n-2} f^k\omega_2(f,f^{n-1-k})
	+O(\hbar^4)
\EEQ
The higher orders of $\hbar$ will not give so simple contributions. The fourth
power, for example, will receive contributions not only from $f^k\omega_4(f,
f^{n-1-k})$ analogous to the $\hbar^2$-terms above, but also from 
$\omega_2(f^k,\omega_2(f^l,f^{n-k-l}))$ and $\omega_2(f^k,f^l)\omega_2(f^m,
f^{n-m-k-l})$.\\
From this it follows that we can write
\BEQA
	{\cal E}_2(f) &=& -\frac{1}{8}\sum_{n=2}^\infty\frac{1}{n!}
	\sum_{k=0}^{n-2} f^k\omega_2(f,f^{n-1-k})\nonumber\\
	&=& -\frac{1}{8}\omega_2(f,f)F_2(f)-\frac{1}{2}\tilde{\omega}_2(f,f)
	G_2(f)
\EEQA
where we have used
\BEQ
	\omega_2(f,f^m) = m f^{m-1}\omega_2(f,f)+m(m-1)f^{m-2}\tilde{\omega}_2
	(f,f)
\EEQ
with
\BEQ
	\tilde{\omega}_2(f,f) := \frac{\partial^2f}{\partial q^2}\left(
	\frac{\partial f}{\partial p}\right)^2-2\frac{\partial^2f}{\partial q
	\partial p}\frac{\partial f}{\partial q}\frac{\partial f}{\partial p}
	+\frac{\partial^2f}{\partial p^2}\left(\frac{\partial f}{\partial q}
	\right)^2
\EEQ
The functions $F_2,G_2$ turn out to be
\BEQA
	F_2(f) &=& \sum_{n=2}^\infty \frac{1}{n!}\sum_{k=0}^{n-2}
	(n-1-k)f^{n-2} = \frac{1}{2}\sum_{n=2}^\infty \frac{1}{n(n-3)!}
	f^{n-2}\\
	G_2(f) &=& \sum_{n=2}^\infty\frac{1}{n!}\sum_{k=0}^{n-2}(n-1-k)
	(n-2-k)f^{n-3} = \frac{1}{6}\sum_{n=2}^\infty \frac{5n-9}{n(n-3)!}
	f^{n-3}
\EEQA
One will be able to write in a similar fashion
\BEQA
	{\cal E}_4(f) &=& \frac{i^4}{2^4 4!}\left(\omega_4(f,f)F_4(f)
	\tilde{\omega}_4(f,f)G_4(f)+\omega_2(f,f)^2F_{2,4}(f)+\right.
	\nonumber\\
	&&\left.\tilde{\omega}_2(f,f)^2G_{2,4}(f)+
	\omega_2(f,f)\tilde{\omega}_2(f,f)H_{2,4}(f)\right)
\EEQA
for suitable functions $F_4,G_4,F_{2,4},G_{2,4},H_{2,4}$. Clearly, as
the power of $\hbar$ grows, these expressions become more and more 
cumbersome. If one would wish to go to arbitrary high powers of $\hbar$
it would be worthwhile to find a diagrammatic expression and some ``Feynman
rules'' for constructing ${\cal E}_n$. It is clear from the recursive
way these are constructed that such ``Feynman rules'' must exist. We will
not attempt to find these rules here, however, but merely restrict ourselves to
noting that often one can assume that powers of $f$ vanish for sufficiently
high powers, or at least become very simple. An example of this could be
$f= \alpha p^2+v(q)$, if the potential is not too big, one can write
$f^n\approx \alpha^n p^{2n}$ for $n$ sufficiently large. Since this particular
example is of great physical importance -- it gives the Hamiltonian of a
Schr\"{o}dinger particle for $\alpha=1/2m$ or the Klein-Gordon field
coupled to some external field in a non-minimal way (e.g. curvature) with
$\alpha=1/2,v(q)=\frac{m^2}{2}+\xi R(q) + V(q)$ -- we will write down the
explicit result for ${\cal E}_2$. From the general formula we see
\BEQ
	{\cal E}_2(\alpha p^2+v(q)) = -\frac{1}{2}\alpha v'' F_2
	-\frac{1}{4}(2\alpha^2 v''p^2+\alpha v^{'2})G_2
\EEQ
The next term, ${\cal E}_4$, will contain third and fourth derivative of
$v(q)$, hence for $v\sim q^k$ for some positive integer $k$ the result
will simplify since the derivatives of $v$ will vanish sooner or later.
For $k=2$, the harmonic oscillator, $v(q)=\frac{1}{2}m\omega q^2, \alpha=1/2m$
we have already
\BEQ
	{\cal E}_2 = -\frac{1}{4}\omega F_2-\frac{1}{8}
	\left(\frac{\omega}{m} p^2+ m\omega^2 q^2\right)
	G_2
\EEQ
We have plotted $e^f+{\cal E}_2(f)$ for $f=p^2+q^2, \hbar=1$ in figure 1.\\
Another interesting case is $v(q)=z/q$ corresponding to the Coulomb
potential (again $\alpha=1/2m$). Here ${\cal E}_2$ becomes
\BEQ
	{\cal E}_2 = -\frac{1}{2m}zq^{-3}F_2-\frac{1}{4m^2}
	\left(zq^{-3}p^2+\frac{m}{2}z^2q^{-4}\right)G_2
\EEQ
and $e^f+{\cal E}_2(f)$ has been plotted for $f=p^2+1/q$ in figure 2.\\
A Yukawa-like potential $v(q) = z\exp(-\mu q)/q$ would lead to
\BEQAS
	{\cal E}_2 &=& -\frac{z}{2m}e^{-\mu q}
	(q^{-3}+\mu q^{-2}+\frac{1}{2}\mu^2q^{-1})
	F_2-\nonumber\\
	&&\frac{z}{8m^2q^4}e^{-2\mu q} \left(2p^2(q+\mu q^2+
	\frac{1}{2}\mu^2q^3) e^{\mu q}+mz(1+2\mu q+\mu^2q^2)\right)G_2
\EEQAS
We have not plotted this, since the exponential factor will only dampen
the behaviour a bit but essentially it would look the same as for $\mu=0$.\\
One should note that for very large values of $n$, the sums defining $F_2,G_2$
will simplify to essentially $\sum_n f^n/n!$, thus
\BEQ
	F_2(-f) = e^{-f}{\cal F}_2(-f) \qquad G_2(-f) = e^{-f}{\cal G}_2(-f)
\EEQ
where ${\cal F}_2,{\cal G}_2$ have slower growth than the exponential,
making $F_2,G_2$ integrable as functions of $p$. This will be needed for the
relationship with the Schwinger-DeWitt expansion to make sense. Moreover,
for $f=\alpha p^2+v(q)$ the integral over $p$ will give $\sigma^{-d/2}$ times
a function of $q$ by simple Gaussian integration. Once more we see that the
factor $\sigma^{-d/2}$ which ``shifts'' the powers of $\sigma$ in an
asymptotic expansion of the heat kernel away from the simple Taylor
series arises in a straightforward manner from the phasespace formalism
upon integrating out the momentum variable.\\
A way to systematically compute such ``quantum corrections'' can be found
by noting that the star exponential is a solution to the differential
equation
\BEQ
	\frac{\partial G}{\partial \sigma} = -f*G
\EEQ
the solution of which is precisely $G=\Exp(-f\sigma)$. Now similarly $g=e^{-f
\sigma}$ is the solution of a differential equation obtained by replacing
the twisted product in the above equation by an ordinary one. Write
\BEQ
	G(q,p;\sigma)=\Exp(-f(q,p)\sigma) = g(q,p;\sigma){\cal G}(q,p;\sigma)
	= e^{-f(q,p)\sigma}{\cal G}(q,p;\sigma)
\EEQ
then we have that $\cal G$ is a solution to
\BEQA
	\frac{\partial{\cal G}}{\partial\sigma} &=& g^{-1} f*(g{\cal G})- f
	{\cal G}\\
	&=&-\frac{1}{2}i\hbar g^{-1}
	\{f,g{\cal G}\}_{\rm PB} + \frac{\hbar^2}{2^2 2!}g^{-1}\omega_2(f,
	g{\cal G})+...
\EEQA
where we have used the definition of the twisted product. Since $g=
e^{-f\sigma}$ the factor $g$ can be pulled out of the Poisson bracket 
cancelling the $g^{-1}$ factor outside. For $f=\alpha p^2+v(q)$ we get
\BEQA
	\omega_2(f,g{\cal G}) &=& 2\alpha g \left[(\sigma^2v^{'2}-
	2 (\sigma-\alpha p^2\sigma^2) v''){\cal G}-\right.\nonumber\\
	&& \left. 2p\sigma v''\frac{\partial {\cal G}}{\partial p}+
	\frac{1}{2\alpha}v''\frac{\partial^2{\cal G}}{\partial p^2}-
	2\sigma v'\frac{\partial{\cal G}}{\partial q}
	+\frac{\partial^2{\cal G}}{\partial q^2}\right]
\EEQA
Supposing $\cal G$ can be Taylor expanded in powers of $\hbar$,
\BEQ
	{\cal G} = \sum_{n=0}^\infty {\cal G}_n\hbar^n
\EEQ
with ${\cal G}_0=1$ we get the following recursive relation
\BEQ
	\frac{\partial}{\partial\sigma}{\cal G}_n = -\sum_{k=1}^n\frac{i^k}{2^k
	k!}g^{-1}\omega_k(f,g{\cal G}_{n-k})
\EEQ
from which we see that ${\cal G}_1=0$ as expected and that
${\cal G}_2$ becomes simply
\BEQ
	{\cal G}_2 = -\frac{1}{8}\int_0^\sigma e^{f\sigma} \omega_2(f,
	e^{-f\sigma})d\sigma
\EEQ
which for the chosen $f$ can be integrated quite readily to give
\BEQ
	{\cal G}_2 = \left(\frac{1}{8}\alpha\sigma^2 v''-
	\frac{1}{12}\alpha\sigma^3(v^{'2}+2\alpha p^2v'')\right) e^{-f\sigma}
\EEQ
By construction we also have
\BEQ
	{\cal G}_2 = e^{f\sigma}{\cal E}_2
\EEQ
leading finally to
\BEQ
	{\cal E}_2 = \frac{1}{12} e^{-v\sigma} \sqrt{\alpha\pi}\sigma^{3/2}
	(2v''-\sigma v^{'2}) 
\EEQ
In a similar way, one can find analytical expressions of the remaining ${\cal 
E}_n$'s.\\
In any case, we can use the expression for the ${\cal E}_{2n}$ to find
expressions for the Schwinger-DeWitt coefficients order by order in
$\hbar$. Examples of this will be given later.\\
For completeness we will quickly list the formula for ${\cal G}_4$.
Straightforward computations yield
\BEQA
	{\cal G}_4 &=& \int_0^\sigma\left(\frac{1}{8}e^{f\sigma}
	\omega_2(f,e^{-f\sigma}{\cal G}_2)
	+\frac{1}{384}e^{f\sigma}\omega_4(f,e^{-f\sigma})\right)d\sigma\\
	&=& \frac{1}{480}\alpha^2\sigma^3 v^{(4)}(5-15\alpha p^2\sigma
	+4\alpha^2p^4\sigma^2)-\frac{1}{288}\alpha^2\sigma^6(v^{'2}
	+2\alpha p^2 v'')^2+\nonumber\\
	&& \frac{1}{240} \alpha^2\sigma^5(9v^{'2}v''+18\alpha p^2 v''+
	4\alpha p^2v'v^{(3)})+\frac{1}{48}\alpha^2\sigma^3v^{(4)}-\nonumber\\
	&&\frac{1}{96}\alpha^2\sigma^2(5v^{``2}+4v'v^{(3)}+\alpha p^2
	v^{(4)})
\EEQA
From this one can then compute ${\cal E}_{2n}$ as
\BEQ
	{\cal E}_{2n} = e^{-f\sigma}{\cal G}_{2n}
\EEQ
As will be clear from these simple examples, the computations are all rather
elementary, allowing one rather quickly to find all the relevant ${\cal 
E}_{2n}$'s, and, we will see, also the coefficients in a Schwinger-DeWitt
asymptotic expansion for the heat kernel.
 
\section{Path Integrals, Wigner Functions and Localisation}
We can make one more important connection. Let $|\zeta\rangle$ be a set of
coherent states (i.e., for the Heisenberg algebra in $n$ dimensions, $h_n$, we
have $\zeta\in\C^n$). We then have
\BEQ
	\Exp(-H_W\sigma) = \int \tilde{\Pi}(\bar{\zeta},\zeta') 
	e^{-\int_0^\sigma \tilde{H}(\bar{\zeta},\zeta') ds}{\cal D}
	(\bar{\zeta},\zeta')
\EEQ
where
\BEQ
	\tilde{O}(\bar{\zeta},\zeta') := \frac{\langle\zeta|O|\zeta'\rangle}
	{\langle\zeta|\zeta'\rangle}
\EEQ
$s$ is a parameter along the path in $\zeta$-space and ${\cal D}(\bar{\zeta},
\zeta')$ is the functional measure
\BEQ
	{\cal D}(\bar{\zeta},\zeta') := \lim_{N\rightarrow\infty}
	\prod_{i=0}^N \frac{\langle\zeta_i|\zeta_{i+1}\rangle}{\langle \zeta_i
	|\zeta_i\rangle}\frac{d\zeta_i}{2\pi i}
\EEQ
with $\zeta_0=\zeta,\zeta_N=\zeta'$. See for instance \cite{path}. This formula
can be read two ways, either the functional integrals allows one to compute
the star exponential, or the star exponential allows one to define a 
regularised functional integral. \\
Introducing the eigenfunctions $|\lambda\rangle$ of $H$, we can write
$H=\sum_\lambda \lambda|\lambda\rangle\langle\lambda|$, but we can also
write the heat kernel as
\BEQ
	G_H=e^{-H\sigma}=
	\sum_\lambda e^{-\lambda\sigma}|\lambda\rangle\langle\lambda|
\EEQ
Thus we arrive at
\BEQA
	\left(e^{-H\sigma}\right)_W &=& \Exp(-H_W(q,p)\sigma)\\
	&=& \sum_\lambda W_\lambda e^{-\lambda\sigma} 
\EEQA
where we have defined
\BEQ
	W_\lambda = (|\lambda\rangle\langle\lambda|)_W
\EEQ
This allows us to ``localise'' a functional integral, turning it into a sum
over a discrete set (the spectrum of the operator). It also says that the
Weyl transform of the heat kernel is a kind of Laplace 
transform ($\lambda\rightarrow\sigma$) of the Wigner function.\\
From the heat kernel one can also compute the Green's function $G=H^{-1}$ by
simply integrating over $\sigma$, 
\BEQ
	G=H^{-1} = - \int_0^\infty e^{-H\sigma} d\sigma
\EEQ
and this implies that we can express the Green's function in terms of the
Wigner function as (using the linearity of the Weyl transform)
\BEQ
	G(q,p) = - \sum_\lambda \lambda^{-1} W_\lambda(q,p) = (H^{-1})_W 
	:= H_W^{*-1}
\EEQ
or
\BEQ
	G(q,p) = \int_0^\infty \Exp(-H_W(q,p)\sigma) d\sigma
\EEQ
giving us an interpretation of the integral of the star exponential.

\section{Applications}
Our first application of the relationship between path-integrals and
star-exponentials involves quantum mechanics. Let
$H=\alpha p^2+v(q)$ be the Hamiltonian for a Schr\"{o}dinger particle in
$d=1$ dimension. Then
\BEQ
	\Pi(u,v) = e^{iu\hat{p}-iv\hat{q}}
\EEQ
is the standard translation operator, giving a ray representation of the
group of translations in two dimensions (i.e. on phasespace), $E_2$. 
We then have
\BEQA
	\Exp(-(\alpha p^2+v(q))\sigma) &=& \int e^{ip\bar{\zeta} -iq\zeta'}
	e^{-\int_0^\sigma \alpha \bar{\zeta}^2+v(\zeta') ds} {\cal D}
	(\bar{\zeta},\zeta')\\
	&=& e^{-(\alpha p^2+v(q))\sigma} + \hbar^2{\cal E}_2(H\sigma) 
	+O(\hbar^4)\\
	&=& \sum_\lambda W_\lambda e^{-\lambda\sigma}
\EEQA
here one can interpret $\sigma$ as an inverse temperature, $\sigma=\beta$. 
In this case, the eigenvalues $\lambda$ are the energies and the $W_\lambda$
become the Wigner functions of the eigenstates, i.e., the solutions to the
time-independent Schr\"{o}dinger equation. For the explicit example of an
harmonic oscillator we then arrive at the relations
\BEQA
	\Exp\left(-\frac{1}{2}(\frac{p^2}{m}+m\omega^2 q^2)\sigma\right) &=&
	\int e^{ip\bar{\zeta}-iq\zeta'}e^{-\int_0^\sigma \frac{\bar{\zeta}^2}
	{2m}+\frac{m\omega}{2}\zeta^{'2} ds}{\cal D}(\bar{\zeta},\zeta')\\
	&=& e^{-(\frac{p^2}{2m}+\frac{1}{2}m\omega^2 q^2)\sigma} 
	-\frac{\hbar^2}{4}
	\omega^2 F_2-\nonumber\\
	&&\frac{\hbar^2}{8}
	\left(\frac{\omega^2}{m} p^2+ m\omega^4 q^2\right)
	G_2+O(\hbar^4)\\
	&=& \sum_{n=0}^\infty W_n(p,q) e^{-\hbar\omega(n+1/2)\sigma}
\EEQA
where
\BEQ
	W_n(p,q) = \frac{(-1)^n}{\hbar\pi n!} 
	e^{-\frac{1}{\hbar\omega}\left(\frac{p^2}{m}+m\omega^2
	q^2\right)}L_n\left(\frac{2}{\hbar\omega}\left(\frac{p^2}{m}+m\omega^2
	q^2\right)\right)
\EEQ
is the Wigner function for the harmonic oscillator, \cite{Dahl}. This also
provides us with an explicit formula for the star exponential and hence of the
path integral. The expression in terms of the Wigner function (i.e., in
terms of Laguerre polynomials $L_n$) show that we can ``localise'' the
functional integral on a countable set, in the same way one can ``localise''
a curve integral in the complex plane on a finite set of poles, the value of 
the integral being proportional to the sum of residues at these poles. Here
we have an integral over a set of more than continuum cardinality (the set of
paths in the complex plane) instead of just a continuum (a specific curve in
the plane), hence the set on which we ``localise'' is no longer finite but
countable. In general one would have the functional integral localised on
a set of at most continuum cardinality (corresponding to a spectrum with
continuous eigenvalues and not just discrete one).\\
The above relation between the star exponential and the Laguerre polynomials
also imply that the Green's function for the harmonic oscillator can be
written as a mode sum
\BEQ
	G(q,p) = \sum_{n=0}^\infty \frac{(-1)^n}{\pi\hbar^2\omega(n+1/2)n!}
	e^{-\frac{1}{\hbar\omega}\left(\frac{p^2}{m}+m\omega^2
	q^2\right)}L_n\left(\frac{2}{\hbar\omega}\left(\frac{p^2}{m}+m\omega^2
	q^2\right)\right)
\EEQ
This is then the phasespace Green's function of this important Hamiltonian.
It turns out that we can find a closed formula for this sum. The generating
function for the Laguerre polynomials is
\BEQS
	\frac{1}{1-t}e^{-\frac{xt}{1-t}} = \sum_{n=0}^\infty \frac{t^n}{n!}
	L_n(x)
\EEQS
Consequently,
\BEQ
	\sum_{n=0}^\infty \frac{(-1)^n}{(n+1/2)n!}L_n(x) = \lim_{t\rightarrow
	-1} t^{-3/2}\int \frac{t^{1/2}}{1-t}e^{-\frac{xt}{1-t}}dt := {\cal L}
	(x)
\EEQ
from which we can see at once that
\BEQA
	G(q,p) &=& \frac{1}{\pi\hbar} {\cal L}(x) e^{-\frac{1}{2}x}\\
	&=& \frac{1}{\pi\hbar}e^{-x/2}\lim_{t\rightarrow -1} t^{-3/2}
	\int \frac{t^{1/2}}{1-t}e^{-\frac{xt}{1-t}}dt
\EEQA
with $x=\frac{2}{\hbar\omega}\left(\frac{p^2}{m}+m\omega^2q^2\right)$. 
Unfortunately, I have not been able to compute this analytically but it 
certainly straightforward to do so numerically.
\\ 
We can also find the Schwinger-DeWitt coefficients $a_n$ in this example.
By performing the integral over $p$ and Taylor expanding the result in
$\sigma$ we get at once
\BEQA
	a_0 &=& \sqrt{2\pi m}+O(\hbar^2)\\
	a_1 &=& -\pi m^2\omega q^2+O(\hbar^2)\\
	a_2 &=& \frac{1}{4}\pi m^3\omega^2 q^4+O(\hbar^2)
\EEQA
and so on.\\
For an arbitrary potential in one dimension we would get
\BEQA
	a_0 &=& \sqrt{2\pi m}+O(\hbar^2)\\
	a_1 &=& -\sqrt{2\pi m} v(q)+O(\hbar^2)\\
	a_2 &=& \frac{\pi m}{2} v^2+O(\hbar^2)
\EEQA
The $\hbar^2$ terms will contain derivatives of $v$ and we see that to lowest
order in $\hbar$ we can treat the potential as a constant. In all cases,
the $O(\hbar^2)$ terms are given by $\int {\cal E}_2dp$.\\
Inserting our expression for ${\cal E}_2$ found above, for $\alpha p^2+v(q)$,
we can readily compute the momentum integral 
\BEQA
	\int {\cal E}_2 dp &=& \frac{1}{12}\sqrt{\alpha\pi}\sigma^{3/2}
	e^{-\sigma v}(2v''-\sigma v^{'2})
\EEQA
Notice that the following powers of $\sigma$ will appear: $3/2,5/2,7/2,....$. 
Thus the power $\sigma^{-1/2}$ appearing in both
the phasespace Taylor series and in the asymptotic Schwinger-DeWitt expansion
comes, as was to be expected, solely from the ``leading symbol'' $p^2$, the
remaining terms only giving higher powers of $\sigma$.\\
The $O(\hbar^2)$ correction, $\delta_2a_n$, to the Schwinger-DeWitt 
coefficients then become
\BEQ
	\delta_2 a_0=\delta_2a_1=0 \qquad \delta_2a_2=\frac{1}{6}\sqrt{\alpha
	\pi}v''
\EEQ
Similarly, by integrating ${\cal E}_4$ one can find the $O(\hbar^4)$
corrections, which will contain even higher powers of $\sigma$.\\
One of the problems with the standard asymptotic expansion is that it ignores
boundary conditions and other global properties of the spacetime manifold and
of the fields. This is a problem, for instance, in computations in curved
spacetimes where one needs some such global contribution (coming from the
vacuum definition) when one wants to find the renormalised energy-momentum
tensor. While definitely not proven, there is some hope that the present
formalism can cure some of these problems, this is so because the expressions
for the heat kernel involves an integral over momentum space, i.e., on a 
curved manifold the cotangent bundle, and this has some information stored
in it about the global properties of the spacetime manifold (how the bundle
is glued together for instance). A full investigation of this problem lies
beyond the scope of the present article which only claims to lay the foundation
and to point out the applicability of the phasespace formalism. \\

As another simple example we will consider a case where the canonical variables
are $\pi_a^i,A^a_i$, with $i=1,2,3$ a spatial index and $a=1,...,n$ a
Lie algebra index, $n=\dim\Lieg$. 
Formally, then, the $\Pi$-function is given by
\BEQ
	\Pi=e^{i\langle u,A\rangle -i\langle v,\pi\rangle}
\EEQ
where $u\in (\Omega^1\otimes\Lieg)^*, v\in\Omega^1\otimes\Lieg$ are
objects ``dual'' to $A,\pi$. The brackets denote the pairing between the
spaces of $\{A,\pi\}$ and their duals. We can thus write
\BEQS
	\langle u,A\rangle = \int u^i_aA^a_i dx 
	\qquad 
	\langle v,\pi\rangle = \int v^a_i\pi^i_a dx
\EEQS
We will be interested in the particular case where
\BEQ
	u^i_a dx = d\gamma^i_a\qquad v^a_idx=d\eta^a_i
\EEQ
where then $\gamma,\eta$ are loops. In this case $\Pi$ becomes
the phasespace analogue of a Wilson loop. As the observable $H$ we will
pick the Chern-Simons functional
\BEQ
	H = -ig\int{\rm tr}(A\wedge dA+\frac{2}{3}A\wedge\ A\wedge A) dx
	=-iS_{\rm CS}[A]
\EEQ
We then have
\BEQ
	\langle\Pi\rangle_t := \int e^{i\oint_\gamma A-i\oint_\eta
	\pi}e^{-i\int_0^tS_{CS}[A]dt}{\cal D}A{\cal D}\pi
\EEQ
This path integral describes a family (indexed by $t$) of Chern-Simons
theories. Writing
\BEQ
	\oint_\eta \pi = \int {\rm Tr}\pi\cdot\dot{\eta} 
	ds = \int {\rm Tr}\pi\cdot\Delta dx
\EEQ
where $\Delta^a_i=\delta(x,\eta(s))\dot{\eta}^a_i$ is
the so-called form factor of the loop, we can perform the $\pi$ integral
obtaining
\BEQ
	\langle\Pi\rangle_t = \delta(\Delta(\eta)) \int 
	e^{i\oint_\gamma A-i\int_0^t S_{\rm CS}[A]dt}{\cal D}A
\EEQ
Two particular cases are interesting, one is where $\{A_t\}$ is a constant
family, i.e., $A_t=A_0\equiv A, \forall t$, the other is where it is a
family centred on a finite number of points, i.e., $A_t\neq 0$ only for
$t=t_0,...,t_N$. In the first case, we have $\int_0^tS_{\rm CS}dt=tS_{\rm CS}$
so putting $k=4\pi gt$ and $q=\exp(\frac{2\pi i}{k+2})$ we have, according
to Witten, \cite{Witten},
\BEQ
	\Exp(-tS_{\rm CS})=\langle\Pi\rangle = 
	\delta(\Delta(\eta)) c(k)^{-w(\gamma)}J_q(\gamma)
	\label{eq:knot}
\EEQ
where $w(\gamma)$ is the writhe of the loop, and $J_q$ is a Jones polynomial.
This also imposes the quantisation condition $k=4\pi gt\in\Z$. The second
case will give a finite sum over such expressions with different coupling
constants $k_t$ and consequently indices $q_t$. We can consider the
general expression as an interpolation between knot invariants for different
values of $k,q$.\\
The equation (\ref{eq:knot}) gives us an interpretation of the star exponential
in this instance, and potentially also another way of computing Jones
polynomials and perhaps of finding relations between them.\\

\section{Conclusion}
We found a relationship between zeta functions, path integrals, star 
exponentials and Wigner functions. This allowed us to regularise path
integrals, derive phasespace expressions for determinants and effective
actions and finally to localise path integrals in the sense of topological
field theory (i.e., write a functional integral as a sum over a set of 
lower cardinality -- either discrete or continuous). 
By Taylor expanding the phasespace expression for the heat kernel, we arrived
at a comparison with the standard Schwinger-DeWitt expansion. We computed the 
Schwinger-DeWitt coefficients of the heat
kernel to $O(\hbar^4)$ explicitly for $f=\alpha p^2+v(q)$. Another explicit
example was taken from topological field theory, where we constructed the
star exponential of the Chern-Simons functional in three dimensions, thereby
getting the Jones polynomials. The application to the heat kernel, however,
seems to be the most promising.

\newpage
\begin{figure}
\caption{The star exponential, $\Exp(f)$ to $O(\hbar^4)$, for $f=p^2+q^2$ in
units with $\hbar=1$.}
\end{figure}

\begin{figure}
\caption{The star exponential for $f=p^2+1/q$ to $O(\hbar^4)$.}
\end{figure}
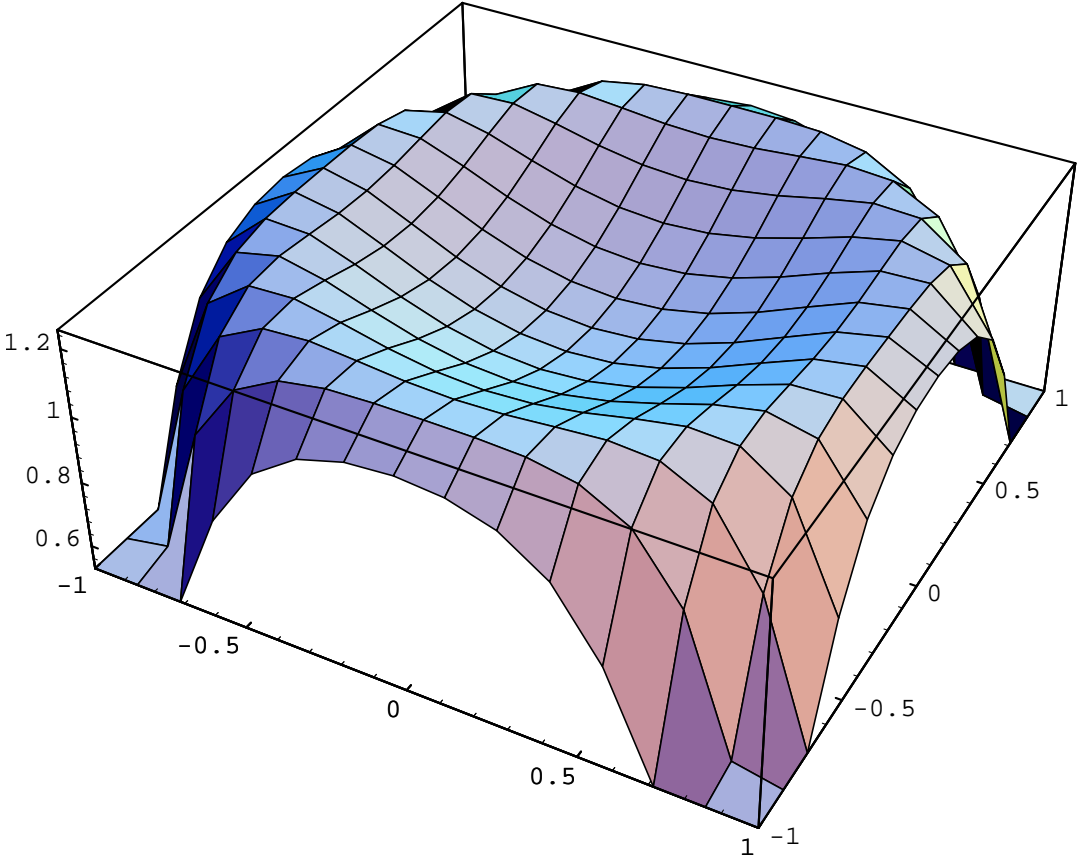
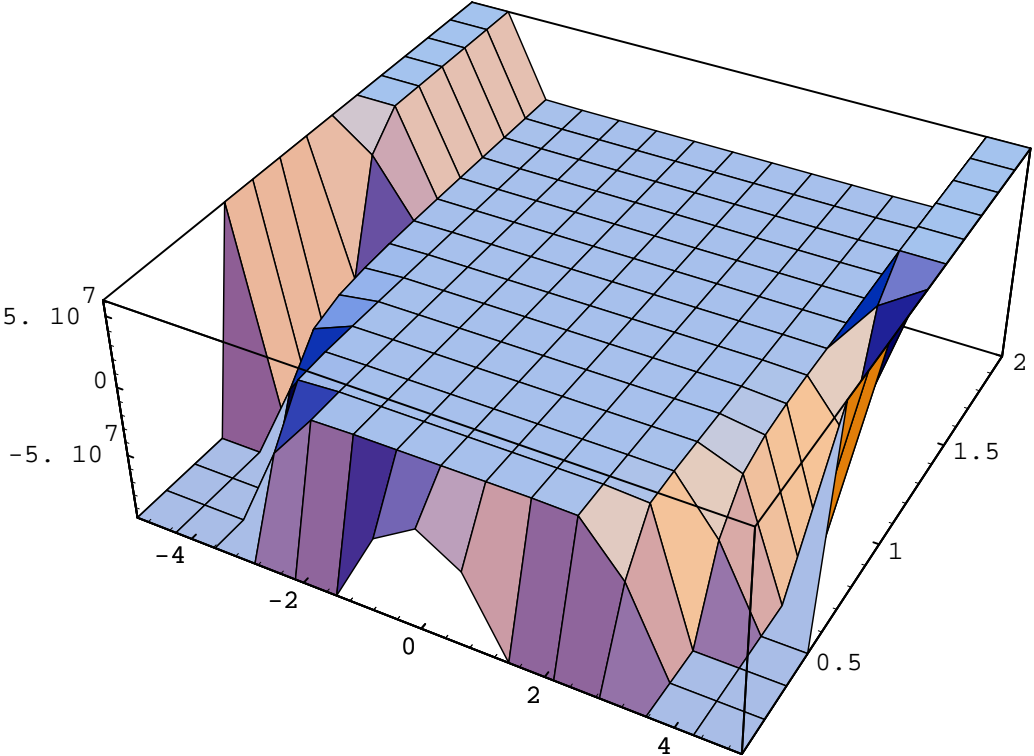
\end{document}